\documentclass[11pt]{article}
\usepackage{moriond,epsfig}
\newcommand{\tz}{\mbox{$\tan \beta \ $}}

\newcommand{\tm}{\mbox{$\tilde m$}}

\newcommand{\bsg}{\mbox{$b \rightarrow s \gamma \ $}}

\newcommand{\dLRbs}{\mbox{$\delta^{LR}_{23} \ $}}
\newcommand{\dLLbs}{\mbox{$\delta^{LL}_{23} \ $}}
\newcommand{\dRLbs}{\mbox{$\delta^{RL}_{23} \ $}}
\newcommand{\dRRbs}{\mbox{$\delta^{RR}_{23} \ $}}
\newcommand{\nn}{\nonumber}
\newcommand{\raw}{\rightarrow}
\newcommand{\lraw}{\leftrightarrow}
\newcommand{\be}{\begin{equation}}
\newcommand{\ee}{\end{equation}}
\newcommand{\bea}{\begin{eqnarray}}
\newcommand{\eea}{\end{eqnarray}}

\begin{document}
\vspace*{4cm}

\title{An alternative approach to \bsg in the unconstrained MSSM}

\author{Stefano Rigolin} 
\address{Theoretical Physics Division, CERN, CH-1211, Geneva 23, 
         Switzerland}

\maketitle\abstracts{
The gluino contributions to the $C'_{7,8}$ Wilson coefficients for 
\bsg are calculated within the unconstrained MSSM. New stringent 
bounds on the \dRLbs and \dRRbs mass insertion parameters are 
obtained in the limit in which the SM and SUSY contributions to 
$C_{7,8}$ approximately cancel. Such a cancellation can plausibly 
appear within several classes of SUSY breaking models. 
Assuming this cancellation takes place, we perform an analysis of the 
\bsg decay. We show that, in the uMSSM 
such an alternative is reasonable and it is possible to saturate the 
\bsg branching ratio and produce a CP asymmetry of up to $20\%$, from 
only the gluino contribution to $C'_{7,8}$ coefficients. Using photon 
polarization a LR asymmetry can be defined that in principle allows  
the $C_{7,8}$ and $C'_{7,8}$ contributions to the \bsg decay to be 
disentangled.
}

%
\section{Introduction}
%
The precision measurements of the inclusive radiative decay $B \raw X_s 
\gamma$ provides an important benchmark for the Standard Model (SM) and 
New Physics (NP) models at the weak scale, such as low-energy 
supersymmetric (SUSY) models. In the SM, flavour changing neutral 
currents (FCNC) are forbidden at tree level. The first SM contribution to 
the \bsg transition appears at one loop level from the CKM flavour 
changing structure, showing the characteristic Cabibbo suppression. 
NP contributions to \bsg typically also arise at one loop, and in 
general can be  much larger than the SM contributions if no mechanisms 
for suppressing the new sources of flavour violation exist (see \cite{all} 
for a complete set of references). 

Experimentally, the inclusive $B \raw X_s \gamma$ Branching Ratio (BR) has 
been measured by ALEPH, BELLE and CLEO, resulting in the current 
experimental weighted average $BR(B\rightarrow X_s \gamma)_{exp} = 
(3.23 \pm 0.41) \times 10^{-4}$,
with new results expected shortly from BABAR and BELLE which could further
reduce the experimental errors. Squeezing the theoretical uncertainties
down to the 10\% level has been (and still is) a crucial task. The SM
theoretical prediction has been the subject of intensive theoretical
investigation in the past several years, leading to the completion of 
the NLO QCD calculations.
The original SM NLO calculation \cite{misiak} gives, for $\sqrt{z} = 
m_c/m_b=0.29$, the following result: $BR(B\rightarrow X_s \gamma)_{SM} = 
(3.28 \pm 0.33) \times 10^{-4}$. The main source of theoretical uncertainty 
is due to NNLO QCD ambiguities. In \cite{gambino} it was shown that using 
$\sqrt{z} = 0.22$ (i.e. the running charm mass instead of the pole mass) 
is more justifiable and causes an enhancement of about 10\% of the \bsg 
BR, leading to the current preferred value: 
$BR(B\rightarrow X_s \gamma)_{SM} = (3.73 \pm 0.30) \times 10^{-4}$. 
Although these theoretical uncertainties can be addressed only with a 
complete NNLO calculation, the SM value for the BR is in 
agreement with the experimental measurement within the $1-2\sigma$ level.

The general agreement between the SM theoretical prediction and the 
experimental results has provided useful guidelines for constraining 
the parameter space of models with NP present at the electroweak 
scale, such as the 2HDM and the minimal supersymmetric standard model 
(MSSM). In SUSY models superpartners and charged Higgs loops contribute 
to \bsg$\!$, with contributions that typically rival or exceed the SM 
one in size. For calculational ease only simplified MSSM scenarios 
(like cMSSM or MSSM with minimal flavour violation (MFV)) have usually been 
assumed. Netherveless, as the origin and dynamical mechanism of SUSY 
breaking are unknown, there is no reason {\it a priori} to expect that the 
soft parameters will be flavour-blind (or violate flavour in the same way 
as the SM). Of course, the kaon system has provided strong FCNC constraints 
for the mixing of the first and second generations, which severely limit 
the possibility of flavour violation in that sector. However the 
constraints for third generation mixings are significantly weaker, with 
\bsg usually providing the most stringent constraints.

A discussion of the \bsg process in the general unconstrained MSSM is in 
principle possible, but it is necessary to deal with two unavoidable 
problems: (i) a large number of free, essentially unconstrained 
parameters; (ii) the need to achieve a quite accurate cancellation 
between the sizeable different contributions (SM, Higgs, chargino, 
neutralino and gluino). 
In the following we'll provide a particularly interesting and simple 
analysis of \bsg in the uMSSM.
%
%
\section{Alternative solution to \bsg branching ratio}
\label{sectiond}
%
%
The low-energy effective Hamiltonian, at the bottom mass scale $\mu_b$, is 
defined as 
${\cal H}_{eff} = - (4 G_F/\sqrt{2}) V_{tb} V^*_{ts} 
   \sum_i C_i(\mu_b) Q_i(\mu_b)$.  
The operators relevant to the \bsg process are: \newline
\parbox{0.5\textwidth}{
\bea 
Q_2 & = & \bar{s}_L \gamma_\mu c_L \bar{c}_L \gamma^\mu b_L \ , \nn \\
Q_7 & = & \frac{e}{16 \pi^2} m_b \bar{s}_L \sigma^{\mu \nu} b_R F_{\mu \nu} \ , 
        \nn \\
Q_8 & = & \frac{g_s}{16\pi^2} m_b \bar{s}_L \sigma^{\mu \nu} G^a_{\mu \nu} 
        T_a b_R \ , \nn
\eea }
\parbox{0.5\textwidth}{
\bea 
Q'_2 & = & \bar{s}_R \gamma_\mu c_R \bar{c}_R \gamma^\mu b_R \ , \nn \\ 
Q'_7 & = &\frac{e}{16\pi^2} m_b \bar{s}_R \sigma^{\mu \nu} b_L F_{\mu \nu} \ , 
         \nn \\ 
Q'_8 & = &\frac{e}{16\pi^2} m_b \bar{s}_R \sigma^{\mu \nu} G^a_{\mu \nu} 
         T_a b_L \ . \label{opQp}
\eea }
Effects of NP generally appear as modifications of the Wilson Coefficients 
(WC) $C^{(_{'})}_{7,8}$ associated to the operators $Q_{7,8}$ and their 
chirality conjugated ones. In the majority of the previous studies of the 
\bsg process, the main focus was to calculate the SM or NP contributions 
to $C_{7,8}$. 
The contributions coming from $C'_{7,8}$ have usually been neglected on the 
assumption that they are suppressed with respect to $C_{7,8}$ by the ratio 
$m_s/m_b$. While this is always valid in the SM, in the 2HDM 
or within specific MSSM scenarios (with MFV), this mass suppression can be 
absent in the uMSSM where the gluino contributions to $C_{7,8}$ and 
$C'_{7,8}$ are naturally of the same order\cite{borzumati}. 

Therefore, in the following we present an alternative approach to \bsg 
in the uMSSM. We assume a particular scenario in which the {\it total} 
contribution to $C_{7,8}$ is negligible and the main contribution to the 
\bsg BR is given by $C'_{7,8}$. This 
``$C'_7$-dominated'' scenario is realized when the chargino, neutralino, 
and gluino contributions to $C_{7,8}$ sum up in such a way as to cancel 
the W and Higgs contributions almost completely. 
In our opinion this situation does not require substantially more fine 
tuning than what is required in the usual MFV scenario, where conversely 
the NP contributions to $C_{7,8}$ essentially cancel between themselves. 

In the following we will focus on the gluino contribution to $C_{7,8}$ and 
$C'_{7,8}$. There is only one gluino diagram that contributes to $C_{7}$ and 
$C'_7$, with the external photon line attached to the down-squark line, while 
two diagrams can contribute to the $C_8$ and $C'_8$ coefficients, as the 
gluon external line can be attached to the squark or the gluino lines. The 
one-loop gluino contributions to the $C'_{7,8}$ coefficients 
are given, at first and second order in the MI, respectively by:
\bea
C^{' \tilde{g}}_7 (1) &=& \frac{8 g_s^2}{3 g^2}{{Q_d}\over{V_{tb} V^*_{ts}}}
   {{m_W^2}\over{\tm^2_D}} \left\{ \delta^{RR}_{23} F^{(1)}_2(x^g_D) - 
   \frac{\tm_{\tilde{g}}}{m_b} \delta^{RL}_{23} F^{(1)}_4(x^g_D) \right\},
\label{glC7MI} \\
C^{' \tilde{g}}_8 (1) &=& - \frac{g_s^2}{3 g^2} {{Q_d}\over{V_{tb} V^*_{ts}}} 
   {{m_W^2}\over{\tm^2_D}} \left\{ \delta^{RR}_{23} F^{(1)}_{21}(x^g_D) - 
  \frac{\tm_{\tilde{g}}}{m_b} \delta^{RL}_{23} F^{(1)}_{43}(x^g_D) 
  \right\}
\label{glC8pMI}
\eea
and 
\bea
C^{' \tilde{g}}_7 (2) &=& \frac{4 g_s^2}{3 g^2}{{Q_d}\over{V_{tb} V^*_{ts}}}
   {{m_W^2}\over{\tm^2_D}} \frac{m_b (A_b - \mu \tz)}{\tm^2_D} 
   \left\{ \delta^{RL}_{23} F^{(2)}_2(x^g_D) - 
   \frac{\tm_{\tilde{g}}}{m_b} \delta^{RR}_{23} F^{(2)}_4(x^g_D) \right\},~
\label{glC7MI2} \\
C^{' \tilde{g}}_8 (2) &=& - \frac{g_s^2}{6 g^2} {{Q_d}\over{V_{tb} V^*_{ts}}} 
   {{m_W^2}\over{\tm^2_D}} \frac{m_b (A_b - \mu \tz)}{\tm^2_D} 
   \left\{ \delta^{RL}_{23} F^{(2)}_{21}(x^g_D) - 
   \frac{\tm_{\tilde{g}}}{m_b} \delta^{RR}_{23} F^{(2)}_{43}(x^g_D) \right\}~. 
\label{glC8pMI2}
\eea
The gluino contributio to $C_{7,8}$ can be obtained exchanging 
L $\lraw$ R in eqs.~(\ref{glC7MI}--\ref{glC8pMI2}).
In deriving eqs.(\ref{glC7MI}--\ref{glC8pMI2}) to the second order in 
the MI parameters, we have kept only the dominant term proportional to \tz 
(the $A_b$ term is retained in the above expression for defining our 
convention for the $\mu$ term; see later) and neglected all of the other 
off-diagonal MIs\footnote{In \cite{all} a complete derivation of the general 
results and conventions used in eqs.~(\ref{glC7MI}-\ref{glC8pMI2}) is 
presented.}. Clearly the dominant terms in eqs.~(\ref{glC7MI}-\ref{glC8pMI2}) 
are those proportional to the 
gluino chirality flip, so that the gluino contribution to $C_7$ ($C'_7$) 
depends, at first order, only on the MI term \dLRbs (\dRLbs$\!$). However, 
for large \tz and $\mu \approx \tm_A$, the second order MI terms in 
eqs.~(\ref{glC7MI2},\ref{glC8pMI2}) can become comparable in size with 
the first order mass insertions. Thus, two different MI parameters are 
relevant in the L/R sectors: (\dLRbs$\!$, \dLLbs$\!$) and (\dRLbs$\!$, 
\dRRbs$\!$), contrary to common wisdom. To which extent the LL and RR MIs 
are relevant depends of course on the values chosen for $\mu$ and $\tan 
\beta$, but in 
a large part of the allowed SUSY parameter space they cannot in general be 
neglected. Moreover, the fact that the gluino WCs depend on two different 
MI parameters will have important consequences in the study of the \bsg 
CP asymmetry.
%
%
\subsection{Single and multiple MI-dominance analysis}
\label{sectionda}
From eqs.(\ref{glC7MI}--\ref{glC8pMI2}), one can read (in MI language) 
the off-diagonal entries that are relevant to the gluino contribution 
to the $C_{7,8}$ and $C'_{7,8}$ WCs. 
Note that limits on \dLRbs $\approx O(10^{-2})$ have previously 
been obtained in \cite{gabbiani}. No stringent bound has been derived 
there for \dLLbs, as this term does not come, at lowest order, with the 
$\tm_{\tilde{g}}/m_b$ enhancement (see eqs.(\ref{glC7MI})). No limits 
were showed on \dRLbs and \dRRbs because the MI formula are symmetric 
in the $L \leftrightarrow R$ exchange and in the scenarios generally adopted 
in the literature the ``opposite chirality'' MIs are suppressed by a factor 
$m_s/m_b$ and so negligible. An analysis of the \dRLbs dependence has 
been performed in \cite{borzumati}, in which the $W$ contribution to 
$C_{7,8}$ was not set to zero (sometimes also Higgs and MFV chargino 
contributions to  $C_{7,8}$ were included). Consequently their bounds on 
the down-squark off-diagonal MIs contributing to $C'_{7,8}$ are 
more stringent than the bounds we derive in our scenario, for which 
the total contribution to $C_{7,8}$ is assumed to be negligible. 
It is clearly only in the scenario we study that an {\em absolute} 
constraint on these MIs can be derived. Moreover no analysis on \dLLbs 
and \dRRbs was performed in \cite{borzumati} as these contributions are 
not relevant in the small \tz region, as can be seen from 
eqs.(\ref{glC7MI2},\ref{glC8pMI2}). 
%
\begin{figure}[t]
\begin{tabular}{cc}
\hspace{-0.5cm}
\epsfig{file=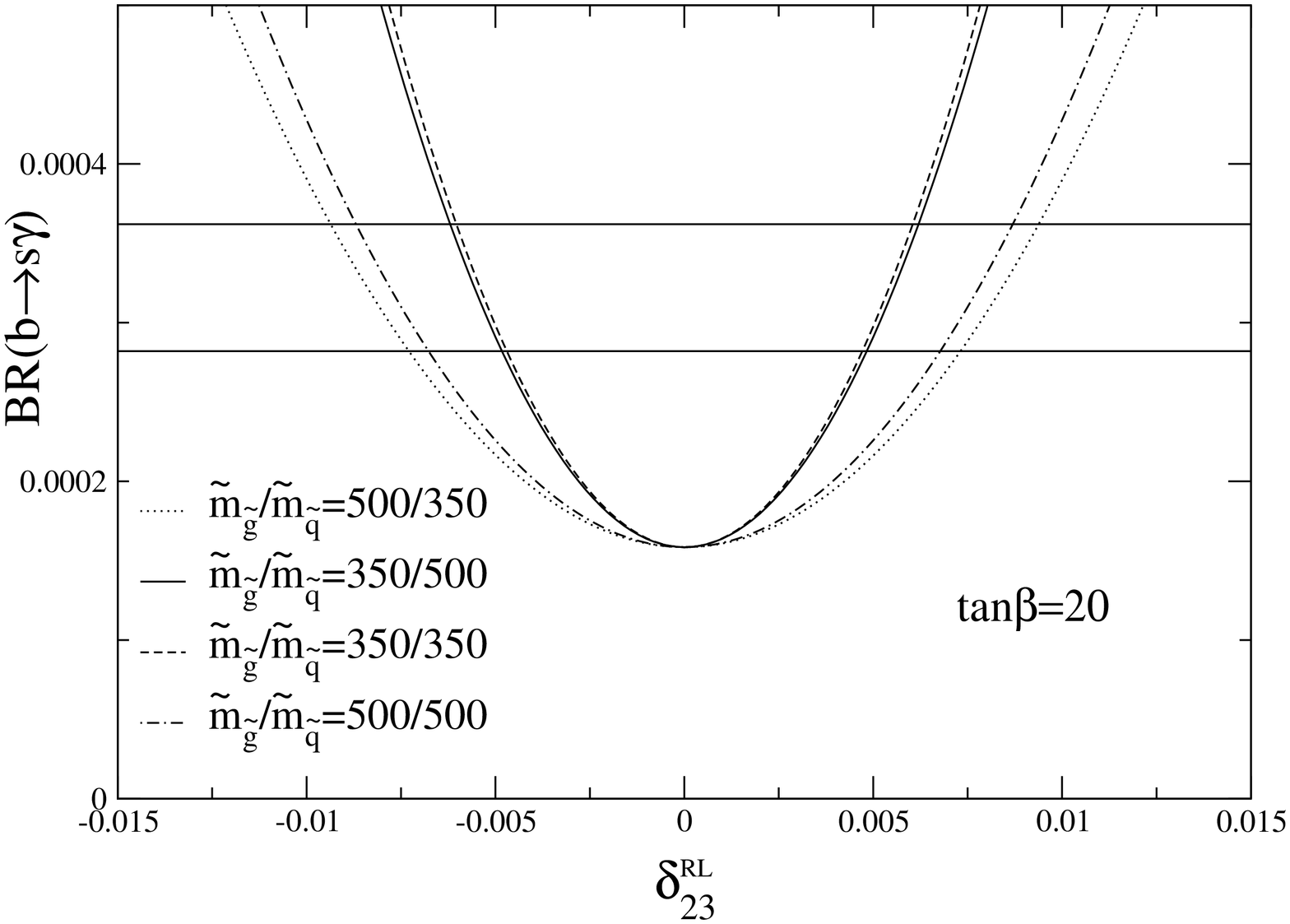,width=6.25cm,angle=-90} &
\hspace{-0.5cm}
\epsfig{file=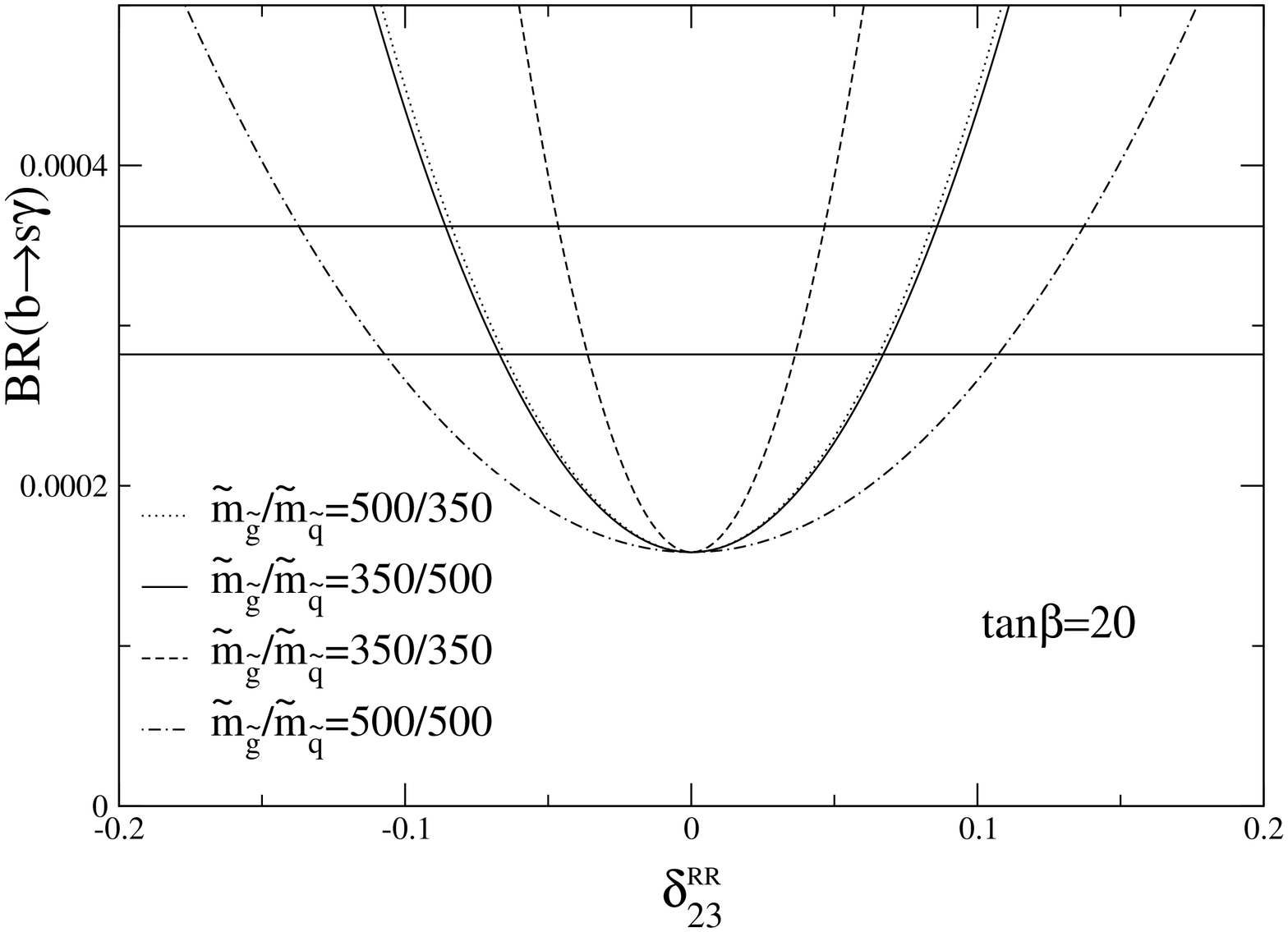,width=6.25cm,angle=-90} 
\end{tabular}
\caption{The dependence of \bsg branching ratio on $\delta^{RL}_{23}$ 
and $\delta^{RR}_{23}$ for different values of $\tm_{\tilde{g}}/\tm_{D}$, 
for $\tz = 20$ and $\mu = 350$ GeV. All of the other off-diagonal entries, 
except the one displayed on the axes, are assumed to vanish. 
$C_{7,8}(\mu_W) = 0$ is assumed. The horizontal lines represent the $1 
\sigma$ experimentally allowed region.}
\label{fig1} 
\end{figure}
%
In fig.~\ref{fig1} we show the dependence of the \bsg BR on the MI terms 
\dRLbs and \dRRbs for different values of $x^g_D=\tm^2_{\tilde{g}}/\tm^2_{D}$ 
and for $\tz = 20$ and $\mu = 350$ GeV. All the other off-diagonal entries 
in the down-squark mass matrix are assumed to vanish. ``Individual'' limits 
\dRLbs $< 10^{-2}$ and \dRRbs $< 1.5 \times 10^{-1}$ can be obtained 
respectively from the left and right side plot of fig.~\ref{fig1}. 
Of course, the required cancellation of the total $C_{7,8}$ contribution 
may in general need non-vanishing 
off-diagonal entries of the up and down squark mass matrices. However, the 
specific values of these entries do not significantly affect the absolute 
limits on the \dRLbs and \dRRbs MIs shown in fig.~\ref{fig1}. As expected 
from eqs.(\ref{glC7MI2},\ref{glC8pMI2}), the bounds obtained 
for \dRRbs are strongly dependent on the product $\mu \tan \beta$. 
More stringent bounds on \dRRbs can be obtained for larger \tz$\!$. For 
\tz$>35$ the bounds on \dRRbs can become as stringent as the \dRLbs bounds. 
%
%
%
%

\begin{figure}[t]
\vspace{0.1cm}
\begin{tabular}{cc}
\hspace{-0.5cm}
\epsfig{file=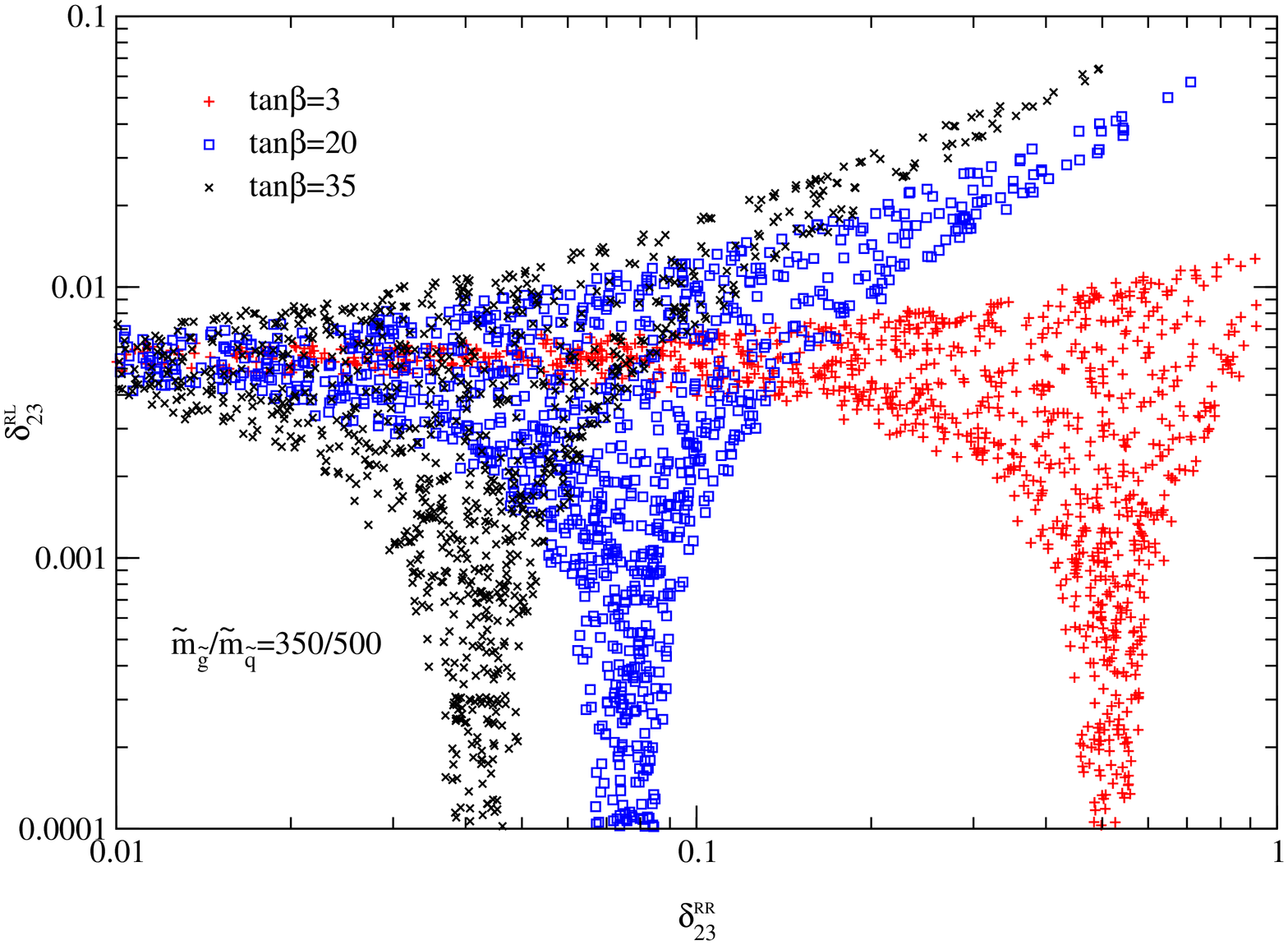, height=8.15cm, angle=-90} & 
\hspace{-0.5cm}
\epsfig{file=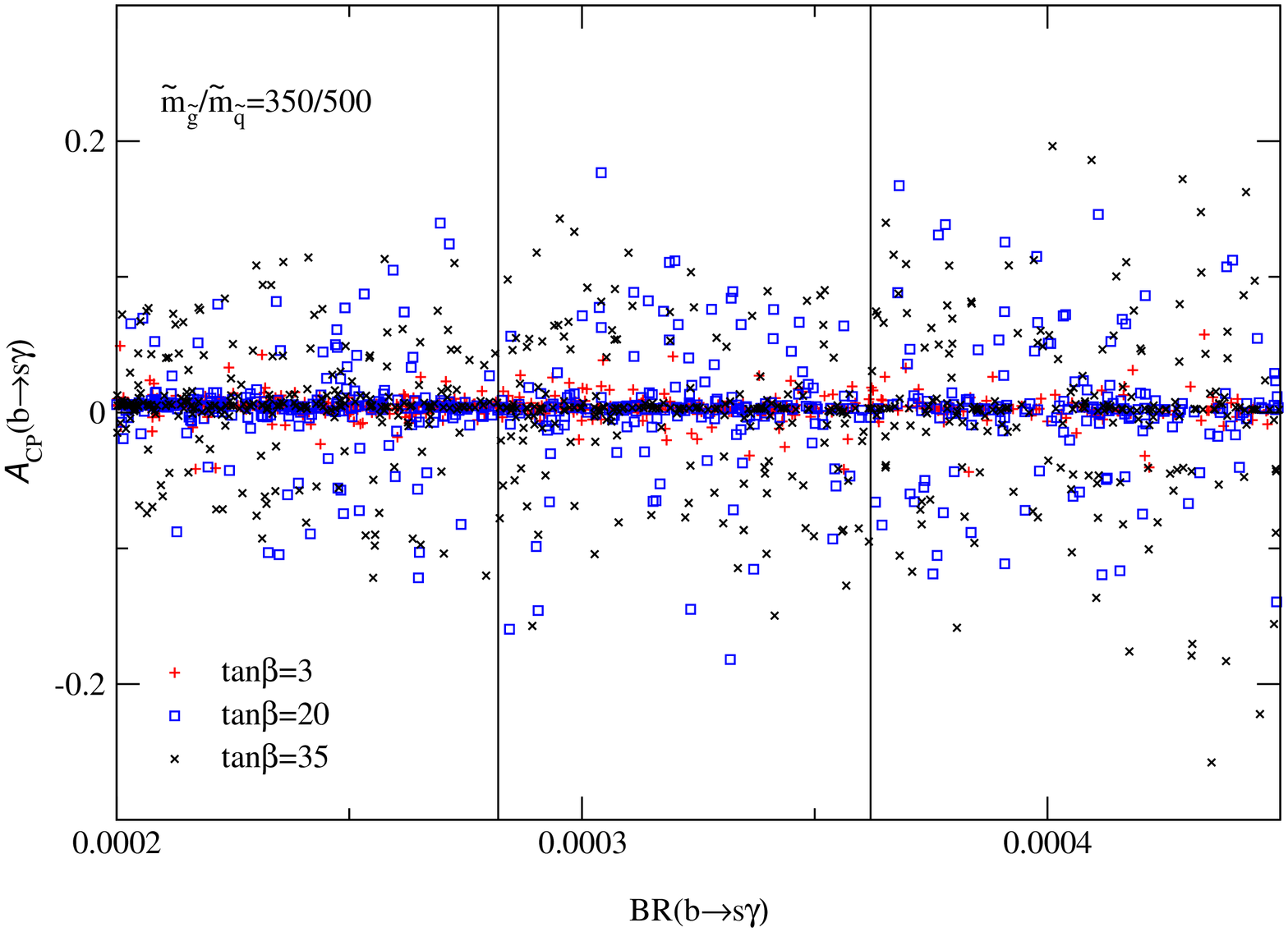, height=8.15cm, angle=-90} 
\end{tabular}
\caption{$1\sigma$-allowed region in the (\dRRbs$\!$, \dRLbs$\!$) 
parameter space (left) and Asymmetry vs Branching Ratio (right) for three 
different values of $\tan \beta$, with the other parameters fixed to 
$\tm_{\tilde{g}}/\tm_{\tilde{q}}=350/500$ and $\mu=350$ GeV. All the other 
off-diagonal entries, except the one displayed on the axes, are assumed 
to vanish.}
\label{fig5} 
\end{figure}

A general analysis of the gluino contribution to $C'_{7,8}$ depends 
simultaneously on both the \dRLbs and \dRRbs MIs. For a complete 
specification of our scenario the only other free parameters that need be 
fixed are the ratio between the gluino mass and the common down-squark mass, 
$\tm_{\tilde{g}}/\tm_D$, the product $\mu$ \tz$\!$, and the relative phase 
between \dRLbs and \dRRbs. The influence of all the other down-sector squark 
matrix off-diagonal entries and MSSM parameters in the $C'_{7,8}$ sector 
can safely be neglected. 
In fig.~\ref{fig5} (left) we show the 
$1 \sigma$ experimentally allowed region in the (\dRLbs$\!$, \dRRbs$\!$) 
parameter space for a specific choice of $\tm_{\tilde{g}}/\tm_D=350/500$, 
$\mu=350$ GeV, and for three different values of \tz=$3, 20$ and $35$. 
For \dRLbs or \dRRbs vanishing, one obtains the regions depicted in 
fig.~\ref{fig1}. 
Larger regions in the (\dRLbs$\!$, \dRRbs$\!$) parameter space are 
obtained when both the MIs take non-vanishing values. It is clear that 
no absolute limit can be derived for the two MIs simultaneously. 
The values (\dRRbs$\!$, \dRRbs$\!$) $\approx (1,0.1)$ are, for example, 
possible 
for \tz$=35$. In fact, as can be seen in fig.~\ref{fig5} (left), there is 
always a ``flat direction'' where large values of \dRLbs and \dRRbs can be 
tuned in such a way that the gluino contribution to $C'_{7,8}$ is consistent 
with the experimental bound. This flat direction clearly depends on the 
chosen values for $\tm_{\tilde{g}}/\tm_{\tilde{q}}$ and $\mu$ \tz$\!$. 
The presence of this particular direction is explained by the fact that 
we are allowing complex off-diagonal entries. Hence the relative phase 
between \dRLbs and \dRRbs can be fixed in such a way that the needed 
amount of cancellation can be obtained between the first and second order 
MI contribution. In the notation used in eqs.(\ref{glC7MI}--\ref{glC8pMI2}) 
the line of maximal cancellation is obtained for 
$\varphi=\arg[$\dRLbs \dRRbs$] =\pm \pi$.  
%

%
\subsection{CP asymmetry and LR asymmetry}
\label{sectiondc}
In addition to the \bsg BR, the experimental collaborations 
will provide in the coming years more precise measurements of the \bsg 
CP asymmetry.
The present experimental value gives, at $90\%$ CL, the range $-0.27 < 
{\cal A}_{CP}(\bsg) < 0.10$ which is still too imprecise to provide useful 
tests for NP, although the measurement is expected to be upgraded soon.

The only flavour-violating and CP-violating source in the SM (and MFV 
scenarios) is given by the CKM matrix, which results in a very small 
prediction for the CP asymmetry. In the SM an asymmetry of approximatively 
$0.5 \%$. If other sources of CP violation are present, a much larger CP 
asymmetry could be produced. In our $C'_7$-dominated scenario, one can derive 
the following approximate relation for the CP asymmetry, in terms of the 
\dRLbs and \dRRbs MIs:
\be
{\cal A}_{CP}(\bsg \!) = -\frac{4}{9}\alpha_s(\mu_b) \frac{ 
   {\rm Im}\left[ C'_7 C'^*_8 \right]}{|C'_7|^2} \approx k(x^g_D)
   \left(\frac{m_b \mu \ \tz}{\tm^2_D} \right) 
   |\dRLbs \dRRbs \!\!| \sin \varphi~,
\label{asymMI}
\ee
where $\varphi$ is the relative phase between \dRLbs and \dRRbs as 
previously defined. 
One can immediately note that if only one MI is 
considered, the CP asymmetry is automatically zero. 
In addition a non-vanishing phase in the off-diagonal down-squark mass 
matrix is necessary. No sensitive bounds on this phase can be extracted 
from EDMs in a general flavour-violating scenario. 


In fig.~\ref{fig5} (right), we show the results obtained for the BR 
and CP asymmetry in which \dRLbs$\!$, \dRRbs and the relative phase 
$\varphi$ are varied arbitrarily for fixed value of $\tm_{\tilde{g}}/
\tm_{\tilde{q}}=350/500$ and \tz$=35$. The full vertical lines 
represent the $1 \sigma$ region experimentally allowed by the 
\bsg BR measurements. It is possible, using $C'_{7,8}$ 
alone, to saturate the \bsg measured BR and at the same 
time have a CP asymmetry even larger than $\pm 10\%$, the sign of the 
asymmetry being determined by the sign of $\sin \varphi$. As shown in 
fig.~\ref{fig5} (right), in the relevant BR range the CP asymmetry 
range is constant. No strong dependence from $\tan \beta$, 
in the large \tz region, is present. The points with large asymmetry 
($>5\%$) lie in the ``flat direction" observed in fig.~\ref{fig5} and 
they have almost $\varphi \approx \pm \pi$ (obviously for $\varphi=\pm 
\pi$ the CP asymmetry vanishes). The explanation of this fact is the 
following. The numerator is proportional to $\sin \varphi$ and so 
goes to 0 as $\varphi$ approaches $\pm \pi$. However, at the same time  
it is enhanced for large MI values. This happens when the flat 
direction condition is (almost) satisfied. Here, in fact, a cancellation 
between the two (large) MI terms takes place, providing the enhancement 
of the CP asymmetry as the denominator remains practically constant, fixed 
by the allowed experimental measurement on the BR. Note also 
that for parameter values outside the flat direction condition a CP 
asymmetry of a few per cent can still be observed, about ten times 
bigger than the SM prediction. 
In our scenario even smaller values of the CP asymmetry can be obtained, 
e.g. if one of the two off-diagonal entries is negligible, or the two MIs 
are ``aligned''. 

%
%
A possible method for disentangling the relative contributions to the 
\bsg BR from the $Q_7$ and $Q'_7$ operators utilizes an   
analysis of the photon polarization. 
For simplicity, let us define the following ``theoretical'' LR asymmetry at LO:
\be
{\cal A}_{LR}(\bsg \! \!) = 
   \frac{BR(b \raw s \gamma_L) - BR(b \raw s \gamma_R)}
        {BR(b \raw s \gamma_L) + BR(b \raw s \gamma_R)} = 
   \frac{|C_7(\mu_b)|^2-|C'_7(\mu_b)|^2}{|C_7(\mu_b)|^2+|C'_7(\mu_b)|^2}~,
\label{asyLR}
\ee
which could in principle disinguish between $C_7$ and $C'_7$ dominated 
scenarios. Here L,R is the polarization of the external photon. This 
quantity is related to the quark chiralities of the $Q_7, Q'_7$ operators. 
Such a measurement is not yet available, as only the average quantity 
$BR(b \raw s \gamma_L) + BR(b \raw s \gamma_R)$ is reported experimentally. 
In the SM case, and in general in all the MFV and mSUGRA scenarios, only 
the $C_7$ coefficient gives a non-negligible contribution to the \bsg 
BR, in such a way that  
${\cal A}_{LR}(\bsg \!\!) = 1$. 
In our scenario, where the total contribution to $C_7$ is negligible, 
one obtains ${\cal A}_{LR}(\bsg\!\!) = -1$. In any other uMSSM scenario, 
any LR asymmetry between $1$ and $-1$ is allowed. Consequently, a 
measurement of $A_{LR}(\bsg \!\!)$ different from one will be a clear 
indication of physics beyond the SM with a non-minimal flavour structure. 
%
%
\section{Conclusions}
%
%
We have discussed an alternative explanation of the \bsg BR in the 
unconstrained MSSM. We analyzed in particular the gluino contribution to 
the WC $C'_7$ associated with the chirality operator $Q'_7$. We show 
that this coefficient arises mainly from two off-diagonal entries: 
\dRLbs and \dRRbs$\!$. For scenarios where the $C_{7,8}$ contributions 
to \bsg are small (i.e. for regions in the MSSM parameter space where 
Ws, Higgs, chargino and gluino contributions to $C_{7,8}$ tend to cancel 
each other), $C'_{7,8}$ provides the dominant effect. We derived absolute 
bounds separately on each of these coefficients. We then described the 
allowed region of (\dRLbs$\!$, \dRRbs$\!$) parameter space, as a function 
of $\tan \beta$. We observed that (for a fixed ratio 
$\tm_{\tilde{g}}/\tm_{\tilde{q}}$ and for each chosen value of $\mu \tan 
\beta$), there exists a ``flat direction'' where large (even $O(1)$) 
off-diagonal entries are allowed. For the majority of parameter space 
the CP asymmetry is less than $5\%$, even if asymmetries 
as large as $20\%$ can be obtained along these ``flat directions''. Finally, 
we suggested that the measure of the LR asymmetry could  
help to disentangle the $C_7$ from the $C'_7$ contribution to the \bsg BR. 
Any ${\cal A}_{LR}(\bsg) \neq 1$ would be an irrefutable proof, not only 
of physics beyond the SM, but also it would indicate the existence of a
non-minimal flavour violation structure of the down-squark mass matrix.
It would be very interesting if such a quantity could be measured. One 
implication of our analysis is that previous results on MSSM parameters, 
including constraints on the ``sign of $\mu$''\footnote{The relative 
sign between the parameters $\mu$ and $A_t$. See \cite{everett} 
for a discussion about the $g-2$ ``sign of $\mu$''.} which are more 
model-dependent than has generally been assumed.

%
%
%
\section*{References}
%
%


\end{document}